\documentclass[aps,prl,twocolumn,superscriptaddress,showpacs]{revtex4}

\usepackage{graphicx}

\begin{document}

%\preprint{}

\title{Low-energy ($< 10$ meV) feature in the nodal electron self-energy and strong temperature dependence of the Fermi velocity in Bi$_{2}$Sr$_{2}$CaCu$_{2}$O$_{8+\delta}$}


\author{N. C. Plumb}
\author{T. J. Reber}
\affiliation{Department of Physics, University of Colorado, Boulder, Colorado 80309-0390, USA}
\email[Electronic address: ]{plumbnc@colorado.edu}
\author{J. D. Koralek}
\affiliation{Lawrence Berkeley National Laboratory, Berkeley, California 94720, USA}
\affiliation{University of California, Berkeley, California 94720, USA}
\author{Z. Sun}
\author{J. F. Douglas}
\affiliation{Department of Physics, University of Colorado, Boulder, Colorado 80309-0390, USA}
\author{Y. Aiura}
\author{K. Oka}
\author{H. Eisaki}
\affiliation{AIST Tsukuba Central 2, 1-1-1 Umezono, Tsukuba, Ibaraki 305-8568, Japan}
\author{D. S. Dessau}
\affiliation{Department of Physics, University of Colorado, Boulder, Colorado 80309-0390, USA}
\affiliation{JILA, University of Colorado and NIST, Boulder, Colorado 80309-0440, USA}
\email[Electronic address: ]{dessau@colorado.edu}

\date{\today}

\begin{abstract}
Using low-photon energy angle-resolved photoemission (ARPES), we study the low-energy dispersion along the nodal $(\pi, \pi)$ direction in Bi$_{2}$Sr$_{2}$CaCu$_{2}$O$_{8+\delta}$ (Bi2212) as a function of temperature.  Less than 10 meV below the Fermi energy, the high-resolution data reveals a novel ``kink''-like feature in the real part of the electron self-energy $\Sigma$ that is distinct from the larger well-known kink roughly 70 meV below $E_F$.  This new kink is strongest below the superconducting critical temperature $T_c$ and weakens substantially as the temperature is raised.  A corollary of this finding is that the Fermi velocity $v_F$, as measured over this energy range, varies rapidly with temperature --- increasing by almost 30\% from 70 to 110 K.
\end{abstract}

%\keywords{}

\maketitle

Angle-resolved photoemission spectroscopy (ARPES) is a direct and powerful probe of electrons and their interactions in solids and is an ideal tool for studying complex materials.  Recently ARPES has accessed a new low-energy photon regime that has substantially improved the technique's resolution, paving the way for several new discoveries \cite{Koralek-2006,Douglas-2007,Iwasawa-2007,Yamasaki-2007,Casey-2008,Ishizaka-2008,Zhang-2008}.  In this Letter we employ low-energy photons to reveal a new interaction effect in Bi$_{2}$Sr$_{2}$CaCu$_{2}$O$_{8+\delta}$ (Bi2212) --- this one located less than 10 meV below $E_F$.  This feature, occurring in the electron self-energy $\Sigma$ along the nodal (gapless) direction of the Fermi surface (FS), has a rapid onset as temperature is lowered past the superconducting critical temperature $T_c$.  Correspondingly, the Fermi velocity $v_F$ has a similar strong temperature dependence.  While electron scattering rate measurements from ARPES have arguably hinted at the existence of this renormalization \cite{Zhang-2008}, this is its first convincing and direct observation.  The new feature may give insights into the physical underpinnings of high temperature superconductivity (HTSC) in layered cuprates.

High resolution has proved crucial to the identification of this new self-energy feature.  As we will show, spectral broadening effects strongly impede attempts to reliably obtain band dispersions near $E_F$ at low temperatures.  Compared to ARPES performed with traditional radiation sources ($h\nu =$ 20--100 eV), low-energy photons ($h\nu =$ 6--7 eV) from lasers and specially configured synchrotron beamlines provide improved momentum resolution and reduced extrinsic spectral broadening due to final state lifetime effects \cite{Koralek-2006}.  The light sources themselves also generally have ultrahigh energy resolution, typically reducing the \emph{total} resolution of the ARPES experiment to well under 10 meV.  Additionally, at these low energies the photoelectrons' mean free path is increased by a factor of 3--10, making ARPES more sensitive to bulk properties \cite{Koralek-2006,Seah-1979}.  As a result, laser- and low energy-ARPES have observed the sharpest spectra from the cuprates to date, representing the best attempts so far at obtaining the intrinsic spectral functions of these materials \cite{Koralek-2006,Casey-2008}.

\begin{figure}[tb]
	\includegraphics[width=0.99\columnwidth]{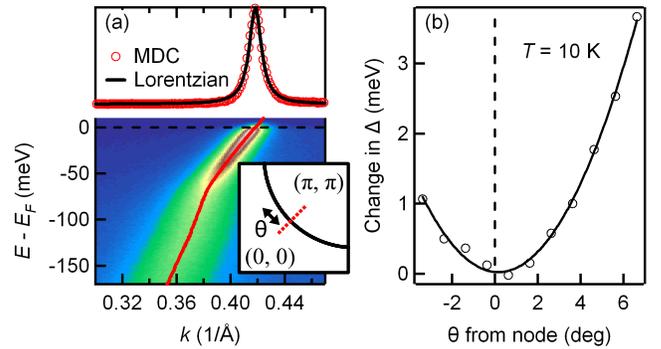} %color
	\caption{\label{fig:fig1}(Color online) Raw ARPES data is shown in (a).  The horizontal dashed line is a single MDC, shown above, which is fit with a Lorentzian.  The Lorentzian peak positions at each energy mark the band dispersion, which is overlaid on the ARPES spectrum.  The inset is a schematic of the ARPES cut (dashed red line) along the nodal $(\pi,\pi)$ direction in the first quadrant of the FS. Nodal alignment is verified by ARPES by scanning the perpendicular angle $\theta$ to study the FS geometry, band velocity, and gap size.  As an example, the minimum of the change in gap size $\Delta$ (b), agrees with the node as determined by FS geometry ($\theta=0$) to high precision (better than $\pm 0.5^{\circ}$).}
\end{figure}

\begin{figure}[tb]
	\includegraphics[width=\columnwidth]{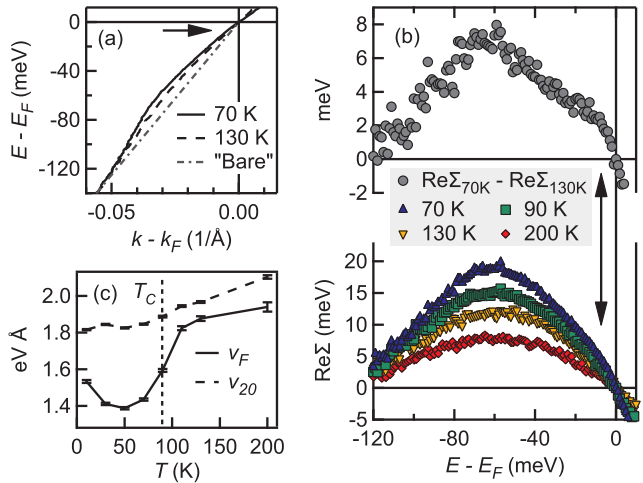} %color
	\caption{\label{fig:fig2}(Color online) MDC-derived dispersions from raw data from nodal Bi2212 (a).  The ``bare band'' (dashed line) is assumed in order to calculate Re$\Sigma$, which is shown as a function of temperature in the lower portion of (b).  In addition to the well-known 70-meV kink renormalization, there is a low-energy (sub-10-meV) feature which is visible below $T_c$ [black arrows in (a) and (b)].  The kink-like nature of the low-energy feature is especially evident in the change in Re$\Sigma$ from 130 K to 70 K [upper portion of (b)]. The nodal Fermi velocity [(c), solid curve], measured via linear fits to the dispersion from -5 to 5 meV, varies strongly with temperature.  The temperature dependence of $v_{20}$ (dashed curve), measured by linear fits from -30 to -10 meV, is far weaker.}
\end{figure}

In the present work, single crystals of optimally-doped Bi2212 ($T_c =$ 91--92 K) were studied using the laser-ARPES system at the University of Colorado  \cite{Koralek-2007} with photon energy $h\nu=$ 6 -- 7 eV, as well as Beamline 5-4 at the Stanford Synchrotron Radiation Laboratory (SSRL) with $h\nu=$ 7 eV.  For the data shown here, total energy resolution $\Delta E$ (photon bandwidth and analyzer resolution) was about 4 meV full-width at half-maximum (FWHM).  Samples were prepared as originally described in \cite{Koralek-2006}.

The complex electron self-energy contains information about all the interactions experienced by an electron.  Here we study Re$\Sigma$, which corresponds to the difference between the observed electron dispersion and the ``bare'' dispersion that would exist if the system were non-interacting.  The analysis process for measuring the band dispersion is illustrated in Fig.~\ref{fig:fig1}.  As has become the standard recently, we utilize momentum distribution curves (MDCs) --– cuts through the ARPES spectrum at constant energies \cite{Valla-1999}.  The Lorentzian peak locations at each energy are taken to mark the dispersion.

Studies of the fine details of the low-energy nodal spectra require unusual alignment accuracy, because the presence of even a small excitation gap can have a profound effect on the near-$E_F$ dispersion. In addition to the usual method using Laue diffraction for azimuthal sample orientation, we took special care to spectroscopically ensure that the dispersions studied were as close to nodal as possible. Using ARPES, we examined the FS geometry, band velocities, and superconducting gap near the node as a function of the angle $\theta$ perpendicular to the ARPES cut [inset of Fig.~\ref{fig:fig1}(a)].  For example, Fig.~\ref{fig:fig1}(b) shows the relative shift in energy of the spectral weight edge (an approximation of the superconducting gap) as a function of $\theta$. The gap minimum agrees with the location of the node determined by the FS geometry to within $\pm 0.5^{\circ}$ ($\pm 0.007$ \AA$^{-1}$ at $h\nu =$ 7 eV).  Based on our studies of the $k$-integrated spectral weights as a function of energy, the maximal gapping of the ``nodal'' spectra is less than 1 meV --– an energy scale insignificant compared to the scale relevant to this work ($E-E_F \lesssim -5$ meV).

Typically MDC fitting results are taken at face value.  While this is acceptable for most energies, in the immediate vicinity of an energetically sharp shift in spectral weight (e.g., near $E_F$), MDC analysis will not yield fully accurate dispersion data if the width of the energy resolution is greater than or comparable to the width of the spectral transition \cite{Ingle-2005}.  For the very fine-scale data near the Fermi energy discussed here, this is an automatic concern and highlights the clear benefit of ultrahigh-resolution low-photon energy ARPES.  With that said, for simplicity's sake, we will first demonstrate the existence and behavior of the sub-10-meV self-energy feature using only raw data. Later on we will turn to the more complicated task of considering the role of instrumental energy resolution.  The analysis will show that when such effects are taken into account, the case for the sub-10-meV kink is even more robust, with the Fermi velocity varying dramatically and essentially monotonically over the full temperature range studied.

Figure~\ref{fig:fig2}(a) illustrates the temperature behavior of the nodal dispersion.  The dashed line connecting two pivot points on the dispersion is taken to be the ``bare'' band, which is subtracted from the measured dispersion in order to calculate Re$\Sigma$.  There is still no agreed-upon method for obtaining the bare band from the data, and the dashed line of Fig.~\ref{fig:fig2}(a) is almost certainly not the true non-interacting dispersion.  However, it is sufficient to take such an \textit{ad hoc} bare dispersion in order to qualitatively assess key features of Re$\Sigma$, as seen in Fig.~\ref{fig:fig2}(b).  The most prominent aspect of the self-energy data presented here is, of course, the well-known ``70-meV kink'' \cite{Bogdanov-2000,Lanzara-2001,Johnson-2001,Kaminski-2001,Gromko-2003}.  In addition to the 70-meV kink, however, a novel bend or kink-like feature is visible at low temperature, located less than 10 meV below $E_F$ (arrow).  This feature is especially evident if one looks at the change in Re$\Sigma$ from 130 to 70 K [shown in the upper portion of Fig.~\ref{fig:fig2}(b)].  

The temperature evolution of the low-energy kink has a direct and dramatic effect on the nodal Fermi velocity.  Figure~\ref{fig:fig2}(c) shows $v_F$ as a function of temperature, where the velocity has been determined by performing a linear fit to each dispersion from 5 meV below $E_F$ to 5 meV above.  (Error bars are plus/minus the standard deviations returned by the fits.)  The velocity $v_{20}$ of the band just below the kink (measured from -30 to -10 meV, relative to $E_{F}$) is also shown.  Compared to $v_{20}$, whose temperature dependence is ostensibly dominated by that of the 70-meV kink, $v_{F}$ varies quite dramatically as a function of temperature.

\begin{figure}[tb]
	\includegraphics[width=\columnwidth]{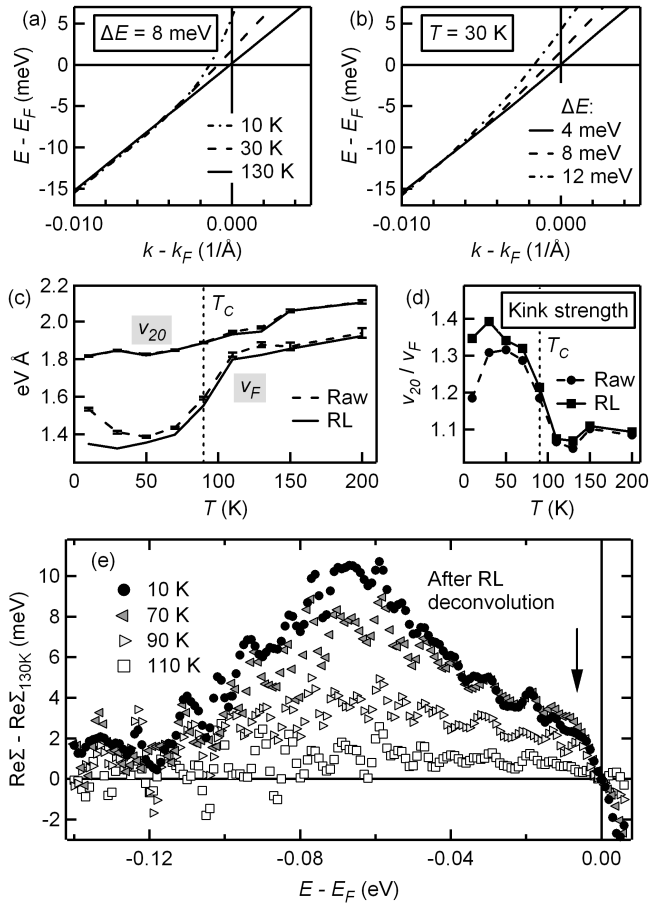} %B/W
	\caption{\label{fig:fig3}Due to the energy resolution of the ARPES experiment, MDC-derived dispersion points near the Fermi energy are deflected to lower momentum.  Simulated ARPES spectra show that this effect strengthens either (a) as the temperature is decreased at a given resolution or (b) as the resolution function is broadened at a given temperature.  Based on 2D RL deconvolution of the spectra, the upturn in $v_F$ seen in the raw data of Fig.~\ref{fig:fig2}(c) below about 50 K is found to be attributable to resolution effects; the Fermi velocity continues to decrease below this temperature (c).  The ``kink strength'' of the low-energy dispersion feature (d) may be quantified as the ratio of the dispersion velocities above and below the feature --- i.e., $v_{20}/v_F$.  By this measure, kink strength increases sharply below $\sim$100 K.  This temperature dependence is reflected in the change in Re$\Sigma$ (of the deconvolved spectra) relative to its value at 130K (e).}
\end{figure}

It is crucial to note that the upturn in $v_F$ below approximately 50 K is an artifact of resolution effects, and in actuality we will show \emph{the Fermi velocity continues to decrease as the temperature is lowered well below 50 K}.  As mentioned previously, at low temperatures and very near $E_F$, energy resolution broadening causes the MDC peaks of a dispersive band to deflect to lower momenta \cite{Ingle-2005} --- the \emph{opposite} direction of the observed kink.  Hence the energy resolution weakens the appearance of this self-energy feature at low temperatures.  The effect is illustrated in Fig.~\ref{fig:fig3}(a) and (b).  Ungapped ARPES spectra were simulated with a linear dispersion through $E_F$ using realistic parameters for nodal data \footnote{The simulations shown use a linear dispersion and a quadratic scattering rate that is roughly consistent with the experimental data.  The findings are not sensitive to the form of the scattering rate (e.g., linear or quadratic).} and then convolved with Gaussian resolution functions.  Lorentzian fits to the resulting simulated MDCs were then performed in the standard way.  The simulations demonstrate that, compared to the intrinsic dispersion, the MDC peaks near $E_F$ will be deflected to lower momentum, ultimately causing the measured Fermi velocity to be greater than the true $v_F$.  The effect becomes more significant as either the temperature is decreased or the resolution function is broadened.  

This non-intuitive result originates simply enough.  EDCs whose peaks lie just below $E_F$ carry significantly more spectral weight than those whose peaks lie just above the Fermi level.  When broadened by convolution with a sufficiently wide energy resolution function ($\Delta E \gtrsim T$), EDCs centered below $E_F$ contribute a disproportionate amount of spectral weight to states above $E_F$.  The redistribution of weight from lower on the band artificially shifts the peaks of MDCs above $E_F$ to lower momenta.  Note that momentum resolution does not have an effect on MDC fitting, since there is no sharp transition in spectral weight along the momentum axis for a constant energy spectrum [although there is one for the energy-integrated spectrum $n(k)$].

To better study the true, underlying behavior of $v_F(T)$, we have removed the effects of energy resolution by performing a two-dimensional (2D) Richardson-Lucy (RL) deconvolution \cite{Richardson-1972,Lucy-1974,Yang-2008}.  We have found this technique to be successful in correcting the dispersion, though at the expense of adding some small ripple.  The technique has been validated based on simulations such as those in Fig.~\ref{fig:fig3}(a) and (b), which agree well with the results of the deconvolution.  The band velocities $v_F$ and $v_{20}$ found from the deconvolved spectra are shown in Fig.~\ref{fig:fig3}(c).  There is little effect on the observed $v_F$ until the temperature drops below about 50 K.  Meanwhile the effect on the deeper band velocity $v_{20}$ is essentially negligible at all temperatures.  The results show that not only is the temperature dependence of $v_F$ quite dramatic in comparison to $v_{20}$ (as was already evident from the raw data), but that it also appears to be monotonic with a strong step in the velocity occurring near $T_c$.  Similarly, the low energy ``kink strength'' defined as the ratio $v_{20}/v_F$, strengthens sharply below $\sim$100 K and possibly saturates somewhat below $\sim$70 K [Fig.~\ref{fig:fig3}(d)].  This behavior is evident in the self-energies of the deconvolved spectra.  Figure~\ref{fig:fig3}(e) shows the change in Re$\Sigma$ from its value at 130 K.  Here again the new low-energy feature (arrow) rises quickly as the temperature drops through 90 K and appears to level off at lower temperatures.  The main 70-meV peak, however, continues to grow as the temperature is lowered down to 10 K.

The origin of the sub-10-meV kink should be of great interest, because the temperature dependence of this feature suggests that it may be associated with superconductivity.  As has been suggested for the 70-meV kink, the sub-10-meV kink could arise from coupling to a bosonic mode (e.g., a magnetic resonance or a phonon) or, as will be discussed, other possibilities may exist as well.

One group has recently claimed that there is an \emph{off-nodal} dispersion renormalization $\sim$20 meV below $E_F$ in La$_{1.85}$Sr$_{0.15}$CuO$_4$ (LSCO) \cite{Sato-2007,Xiao-2007}, which the authors speculate is associated with an incommensurate magnetic resonance near 9 meV seen in inelastic neutron scattering (INS) data \cite{Tranquada-2004a}.  However, rather than being an isolated, sharp mode, the 9-meV resonance may belong to an incommensurate magnetic resonance branch that disperses over a broad energy range \cite{Tranquada-2004b}, and hence it is not clear that the well-defined low-energy renormalization along the node should be attributed to a magnetic resonance.

Theoretical calculations have routinely predicted a sizable, distinct peak in the phonon density of states near or below 10 meV in cuprate superconductors \cite{Kress-1988,Stachiotti-1995,Giustino-2008}. Moreover, various experimental studies of the cuprates have observed phonons near or below 10 meV attributable to $c$-axis longitudinal and transverse optical modes \cite{Tsvetkov-1999}, as well as in-plane transverse optical and acoustic phonons with very low-lying dispersions \cite{Gauzzi-2007,dAstuto-2003}.  However, if phonons are responsible for the renormalization, this implies scattering between near-nodal electron states, since most of the states away from the node at the relevant energy scale ($<10$ meV below $E_F$) are eliminated below $T_c$ due to the $d$-wave superconducting gap.  This contrasts with the 70-meV kink in the self-energy, which lies beyond the gap energy, and therefore, if due to electron-phonon interactions, may couple states over a broad momentum space.  Whether or not electron-phonon coupling at either energy might support or oppose superconductivity remains to be seen.

It is worthwhile to consider alternatives to the electron-boson coupling paradigm. Chubukov and Eremin have recently claimed that the observed temperature scaling of $v_F$ is more-or-less consistent with non-analytic corrections to Fermi liquid theory in 2D \cite{Chubukov-2008}.  They claim that the form and size of the low-energy feature are roughly what is expected from the theory.

The observed sub-10-meV feature might also be consistent with the theory of a marginal Fermi liquid (MFL) \cite{Varma-1989} applied to the superconducting state. An MFL is characterized by Im$\Sigma \propto$ max$(|\omega|, T)$.  The discontinuity at $|\omega| = T$ is merely an ansatz, and in reality one should expect a smooth crossover in the behavior of Im$\Sigma$ at some very low energy related to $T$.  The Kramers-Kronig transformation of such a crossover might give rise to a low-energy kink similar to what has been observed here.  The data suggests that the location of the kink does not vary with temperature, which argues against the MFL theory.  However, given the subtlety of the feature and the difficulty of extracting it from the data, it is hard to definitively rule out MFL behavior.

In conclusion, a new electron self-energy feature has been observed at extremely low energy along the superconducting node in Bi2212.  This kink-like feature lies less than 10 meV below $E_F$. It is visible at temperatures below $T_c$ and weakens rapidly at higher temperatures.  Associated with this feature, the Fermi velocity scales substantially --- increasing by roughly 30\% from 70 to 110 K.  The temperature dependence of the feature suggests a possible role in superconductivity, although it is unclear at this time what mechanism(s) may lead to this low-energy renormalization.

\begin{acknowledgments}
Funding for this research was provided by DOE Grant No. DE-FG02-03ER46066 with partial support from the NSF EUV ERC.  We thank A. V. Chubukov, T. P. Devereaux, S. Johnston, and K. Shimada for valuable conversations.  We are deeply grateful to Q. Wang, J. Griffith, S. Cundiff, H. Kapteyn, and M. Murnane for assistance. We also thank D. H. Lu and R. G. Moore for assistance at SSRL. SSRL is operated by the DOE, Office of Basic Energy Sciences.
\end{acknowledgments}

\bibliography{citations}

\end{document}